%% file: main.tex
\documentclass[runningheads]{llncs}

\usepackage[T1]{fontenc}

\usepackage{graphicx}
\usepackage{multirow}
\usepackage{stfloats}
\usepackage{xcolor}
\usepackage{url}
\usepackage{booktabs}

\usepackage[table]{xcolor}
\usepackage{tikz}
\usepackage{diagbox}
\usetikzlibrary{positioning,fit,shapes}

\usepackage{tablefootnote}
\usepackage{listings}
\usepackage[edges]{forest}
\usepackage[most]{tcolorbox}

\usepackage{color}
\usepackage{nameref}
\usepackage{tabularx}
\usepackage{censor}
\usepackage{mdframed}
\usepackage{fontawesome5}

\definecolor{lightblue}{RGB}{0,0,100}

\definecolor{purplish}{HTML}{D8DFE3}
\definecolor{purplishlight}{HTML}{EBEFF3}
\definecolor{purplishdark}{HTML}{007ACC}
\definecolor{myorange}{HTML}{007ACC}

\definecolor{lightpurple}{RGB}{200,229,204}
\definecolor{darkpurple}{RGB}{200,140,0}

\newtcolorbox{MyBox}{
  colback=white,
  colframe=lightblue,
  fonttitle=\bfseries,
  coltitle=black,
  sharp corners,
  boxrule=1pt,
  left=5pt,
  right=5pt,
  top=5pt,
  bottom=5pt,
  breakable
}

\newmdenv[
  backgroundcolor=yellow!8,
  leftline=true,
  rightline=false,
  topline=false,
  bottomline=false,
  linecolor=black,
  linewidth=3pt,
  innerleftmargin=10pt,
  innerrightmargin=10pt,
  innertopmargin=8pt,
  innerbottommargin=8pt,
  skipabove=8pt,
  skipbelow=8pt
]{resultbox}

\newtcolorbox[auto counter,number within=section]{rqbox}[2]{
    nameref=#1,
    title=\small{#1},
    enhanced,
    attach boxed title to top left={yshift=-6pt,xshift=8pt},
    boxed title style={size=small,boxsep=1pt},
    colframe=purplishdark,
    colback=white,
    colbacktitle=purplishdark,
    boxsep=2pt,
    left=2pt,
    right=2pt,
    top=6pt,
    bottom=2pt,
    middle=2pt
}

\newtcolorbox[auto counter,number within=section]{promptbox}[2]{
    nameref=#1,
    title=\small{#1},
    enhanced,
    attach boxed title to top left={yshift=-6pt,xshift=8pt},
    boxed title style={size=small,boxsep=1pt},
    colframe=myorange,
    colback=white,
    colbacktitle=myorange,
    boxsep=2pt,
    left=2pt,
    right=2pt,
    top=6pt,
    bottom=2pt,
    middle=2pt
}

\newmdenv[
  backgroundcolor=lightpurple,
  linecolor=darkpurple,
  linewidth=3pt,
  leftline=true,
  topline=false,
  bottomline=false,
  rightline=false,
  innerleftmargin=10pt,
  innerrightmargin=10pt,
  innertopmargin=5pt,
  innerbottommargin=5pt,
  skipabove=10pt,
  skipbelow=10pt
]{participantbox}

\begin{document}

\title{SLRMentor: An LLM-Based Tool Supporting Learning of SLR in Software Engineering}

\titlerunning{SLRMentor}

\author{
Rodolfo Gil-Pereira\inst{1}
\and
Ronnie de Souza Santos\inst{1}
\and
Italo Santos\inst{2}
\and
Cleyton Magalhães\inst{3}
}

\authorrunning{R. Gil-Pereira et al.}

\institute{
University of Calgary, Canada
\and
University of Hawai‘i at Mānoa, USA
\and
Universidade Federal Rural de Pernambuco (UFRPE), Brazil
\\
\email{
rodolfo.gilpereira@ucalgary.ca,
ronnie.desouzasantos@ucalgary.ca,
isantos3@hawaii.edu,
cleyton.vanut@ufrpe.br}
}

\maketitle

\begin{abstract}

This paper presents \textit{SLRMentor}, a conversational assistant designed to support both learning about the systematic literature review process and the execution of planning activities in software engineering. The tool offers general guidance on SLR methodology and supports key planning tasks, including search string construction and reasoning about inclusion and exclusion criteria, with explanations grounded in established SLR guidelines. A pilot validation with graduate students suggests that \textit{SLRMentor} helps clarify the SLR process and planning decisions, lowers initial barriers for novice researchers, and supports learning while still requiring active methodological judgment.


\end{abstract}

\input{intro}

\input{back}
\input{method}
\input{results}

\input{discussion}
\input{conclusions}

\bibliographystyle{splncs04}
\bibliography{bib}

\end{document}

%% file: intro.tex
\section{Introduction}
\label{sec:introduction}

Evidence-based software engineering (EBSE) advocates the systematic use of empirical evidence to inform software engineering research and practice \cite{kitchenham2004ebse}. Within EBSE, systematic literature reviews (SLRs) are defined as secondary studies that identify, evaluate, and synthesize primary research through explicit, structured, and repeatable procedures \cite{kitchenham2004ebse,kitchenham2009slr,ralph2020empiricalstandards}. SLRs are widely used in software engineering to consolidate empirical evidence, reduce reliance on anecdotal knowledge, and support theory building and research agenda setting, particularly in a field characterized by diverse study contexts, research methods, and terminology \cite{kitchenham2009slr,petersen2008sms,zhang2010searching}. Conducting an SLR involves a sequence of well-defined activities, including planning the review, executing the search, selecting relevant studies, extracting data, and synthesizing findings \cite{kitchenham2009slr,petersen2008sms,ralph2020empiricalstandards}. Among these activities, the planning phase is consistently identified as central to review quality and credibility, as it determines the research questions, inclusion and exclusion criteria, and search strategies that guide subsequent steps \cite{zhang2010searching,wohlin2020updatesearch}. Prior work reports that search strategy construction is time-consuming and error prone due to terminology variation, database specific syntax, and trade-offs between recall and precision, and that planning decisions directly affect the completeness, reproducibility, and updatability of SLRs \cite{zhang2010searching,kitchenham2009slr,wohlin2020updatesearch}.

Recent research has investigated the use of artificial intelligence techniques to support specific activities within the SLR process \cite{bolanos2024ailitreviews,delatorre2023ai}. Existing studies report that large language models (LLMs) have been explored primarily for supporting search related tasks and study selection activities, including keyword suggestion and classification based on titles and abstracts \cite{bolanos2024ailitreviews,felizardo2024chatgptslr}. Empirical evaluations indicate that such AI-based support exhibits limitations in accuracy and reliability and should therefore be used as an assistive aid rather than as a substitute for human judgment \cite{felizardo2024chatgptslr}. The reviewed literature further shows that current AI based approaches concentrate on improving efficiency in executing review tasks, particularly during study selection, while providing limited support for methodological understanding or learning of the SLR process itself \cite{bolanos2024ailitreviews,felizardo2024chatgptslr}.

Despite established guidelines and the increasing availability of tool support, prior educational research indicates that teaching evidence-based software engineering and systematic literature review methods remains challenging, particularly for novice researchers \cite{castelluccia2013teaching,riaz2010experiences,santos2013motivation}. Studies report that effective learning of systematic reviews requires more than procedural guidance, with students benefiting from instructional support that makes methodological reasoning and decision making explicit and supports iterative engagement across planning and protocol development activities \cite{castelluccia2013teaching,riaz2010experiences,pizard2021training,santos2014smsreviews}. Motivated by these observations, this research investigates how a conversational assistant can support learning of SLR planning activities in software engineering and poses the following research question (RQ):

\begin{resultbox}
\textbf{RQ. \textit{How can a conversational assistant support researchers in learning how to conduct systematic literature reviews in software engineering?}}
\end{resultbox}

To address this question, we present \textit{SLR Mentor}, a LLM-based conversational assistant designed to support learning of SLR practices with a focus on planning activities, providing explanations, and structured guidance grounded in established SLR guidelines~\cite{kitchenham2009slr}.

%% file: back.tex
\section{Background}
\label{sec:background}

This section provides background on two strands of prior work that inform this study: research on the teaching and learning of systematic literature reviews in software engineering and the use of conversational agents in educational contexts.

\paragraph{Teaching Systematic Literature Reviews in Software Engineering.} Research on evidence-based software engineering reports that systematic literature reviews and systematic mapping studies are commonly taught in graduate and advanced undergraduate software engineering courses \cite{castelluccia2013teaching,jorgensen2005teaching}. Instruction typically combines theoretical introductions to EBSE principles with project-based activities in which students conduct secondary studies on selected topics, often grounded in established SLR guidelines \cite{castelluccia2013teaching,santos2014smsreviews}. Prior work indicates an emphasis on hands-on engagement with the review process, supported by problem-based and project-based learning approaches that connect methodological concepts to concrete research questions and domains \cite{castelluccia2013teaching}. The literature also indicates that learning to conduct SLRs presents challenges for novice researchers across multiple stages of the process, including research question formulation, search strategy construction, inclusion and exclusion criteria definition, quality assessment, and synthesis \cite{riaz2010experiences}. These difficulties have been attributed to limited experience with secondary research methods and domain knowledge, and may persist even when guidelines are followed \cite{riaz2010experiences}. Educational studies suggest that without explicit instructional support, students may focus on producing required artifacts rather than engaging with underlying methodological reasoning, motivating calls for teaching approaches that support reflection, iteration, and explicit discussion of planning decisions \cite{castelluccia2013teaching,pizard2021training,santos2014smsreviews}.

\paragraph{Conversational Agents for Learning Support in Software Engineering.} Conversational agents and chatbots are commonly described as software systems that interact with users through natural language to provide information, guidance, or automated assistance \cite{cunningham2019review,wessel2022bots}. In educational contexts, they are characterized as interfaces that mediate access to learning resources, support engagement through dialogue-based interaction, and assist learners during instructional activities, including responding to inquiries, supporting self-paced learning, and providing feedback \cite{cunningham2019review,chan2023potential}. Prior research also reports limitations related to transparency, consistency of responses, and alignment with pedagogical goals \cite{cunningham2019review,chan2023potential}. More recent work indicates that conversational agents are increasingly implemented using LLMs, expanding expressive capacity while raising concerns related to reliability, student reliance, instructional control, and academic integrity \cite{khan2025integrating}. Within software engineering education, these systems have been investigated as support mechanisms for programming practice, guidance on software engineering processes, and interaction with educational or development platforms, with studies reporting potential value for explanation and assistance alongside concerns about variability in learning experiences and an emphasis on task completion over conceptual understanding \cite{wessel2022bots,sengul2024software,fernandez2024exploring,khan2025integrating}. Across this literature, conversational agents are reported as complementary tools whose role depends on their integration with instructional design and learning objectives rather than as substitutes for teaching or learning activities \cite{wessel2022bots,khan2025integrating}.

%% file: method.tex
\section{Method} \label{sec:method}

This study follows the engineering research method, also referred to as Design Science~\cite{ralph2020empiricalstandards}. Our contribution is a software artifact, \textit{SLRMentor}, designed to support novice software engineering researchers in planning systematic literature reviews. The tool provides learning-oriented guidance during the early stages of SLRs, with a focus on supporting understanding and methodological reasoning rather than automating review tasks. \\

\noindent\textbf{Tool Development.} The development of \textit{SLRMentor} was informed by prior work on conversational agents and bots in software engineering and educational contexts, which characterizes such systems as interfaces for providing guidance, information, and learning support through natural language interaction \cite{wessel2022bots}. The tool employs a natural language conversational interface to support incremental, dialogue-based interaction during SLR planning, reflecting how conversational agents have been used to assist users in complex tasks and learning activities \cite{wessel2022bots}. Two general-purpose LLMs (OpenAI and Google Gemini) are used to respond to general questions about the SLR method and to support specific planning activities, such as the construction of search strings and the definition of inclusion and exclusion criteria. Generated outputs are accompanied by explanations intended to support methodological reasoning rather than direct adoption, consistent with prior work emphasizing the assistive rather than autonomous use of such systems in educational settings \cite{fernandez2024exploring,khan2025integrating}. The system supports iterative refinement by allowing users to revise goals and artifacts across multiple interactions, aligning with descriptions of how SLR protocols are developed and refined in practice \cite{kitchenham2009slr}. To support reliability, \textit{SLR Mentor} integrates retrieval-augmented generation \cite{swacha2025retrieval,fan2024survey}, grounding responses in curated SLR guidelines and methodological documentation to improve consistency with established practices and increase transparency, including but not limited to \cite{petersen2008sms,ralph2020empiricalstandards}. \\

\noindent\textbf{Task Definition.} The functionality of \textit{SLRMentor} is structured around three learning oriented tasks aligned with the planning phase of systematic literature reviews. First, the Mentor Chat supports conceptual understanding of the SLR process by allowing users to ask questions about review stages, methodological considerations, documentation practices, and common sources of bias, with the aim of supporting orientation and sense-making, particularly for novice reviewers. Second, the Search String Chat assists users in constructing search strategies by guiding the translation of a research goal into a structured search string, proposing keywords and related terms, explaining Boolean operators, and encouraging reflection on alternative formulations, with an emphasis on understanding systematic search design rather than simply producing a final string. Third, the Criteria Chat supports the definition of inclusion and exclusion criteria by helping users derive criteria from their study goals and explaining how these criteria operationalize scope and relevance, thereby supporting reasoning about study selection decisions and their implications for the review. \\

\noindent\textbf{Validation.} The pilot validation of \textit{SLRMentor} adopted an exploratory and formative approach focused on educational value and perceived reliability when used by novice researchers, consistent with the early stage of the artifact and an Engineering Research perspective \cite{ralph2020empiricalstandards}. The evaluation took place in the context of a graduate software engineering course with eight students, comprising four doctoral and four master’s students, whose final course assignment required them to conduct either a mapping study or a rapid review, both of which are types of systematic literature reviews, on their own research topics following established SLR guidelines. All students were novice reviewers, conducting a secondary study for the first or second time. After course completion and grading, students were invited to voluntarily and anonymously use \textit{SLR Mentor} and provide feedback through a structured questionnaire, with no impact on course assessment; four students chose to participate. Participants were instructed to interact with all three components of the system and to apply them to the same study they had previously conducted manually, enabling reflection and comparison between manual and tool-supported planning. Data collection combined Likert scale items and open-ended questions addressing clarity, adequacy, consistency with known SLR practices, and perceived support for learning and reflection. Analysis emphasized descriptive summaries and qualitative interpretation rather than statistical inference, with reliability considered indirectly through participants’ comparisons between tool-generated artifacts and their manually produced counterparts, focusing on perceived methodological alignment, transparency, and support for critical reasoning. The validation data is available at \url{https://figshare.com/s/e332968f559de828cbab}.

%% file: results.tex
\section{Results}
\label{sec:findings}

\begin{figure*}[h]
\centering
\includegraphics[width=1\linewidth]{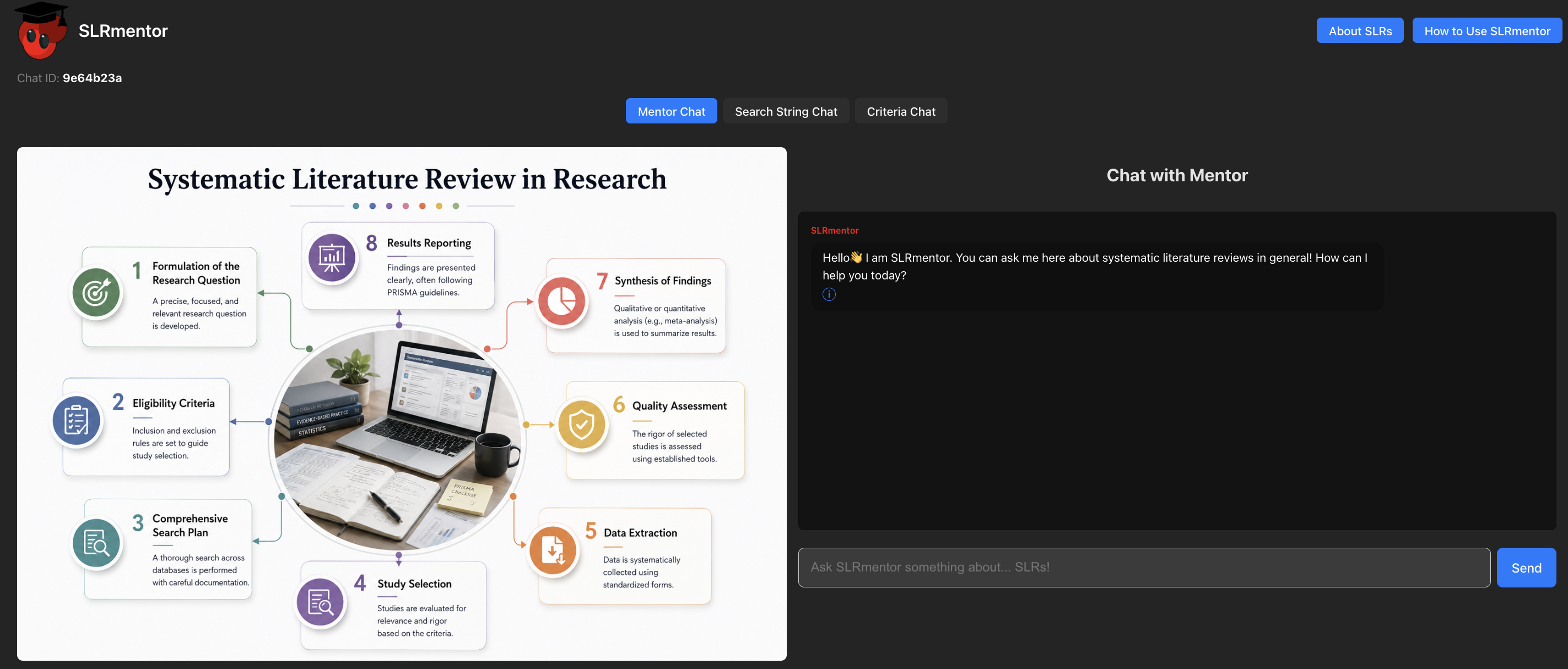}
\caption{SLRMentor Interactive Interface in Use}
\label{fig:tool}
\end{figure*}

In this paper, we focus on the educational role of \textit{SLRMentor}, describing how its features can be used to support SRL teaching and learning and how novice researchers experienced these features. Figure \ref{fig:tool} shows an overview of \textit{SLRMentor}, and the live version is available at \url{https://slrmentor.ca/} \\

\noindent \textbf{Educational Use of SLRMentor.} \textit{SLR Mentor} supports learning during the general understanding and planning phases of systematic literature reviews through three complementary components:

\begin{itemize}
    \item \textbf{Mentor Chat.} The Mentor Chat provides general guidance on systematic literature reviews, supporting conceptual understanding of review stages, methodological principles, and common decisions involved in conducting an SLR. This component is intended to help students clarify doubts about the review process and reflect on methodological choices.

    \item \textbf{Search String Chat.} The Search String Chat supports the construction of search strings by guiding students in translating a research goal or question into searchable terms. This feature provides explanations about keyword selection, use of synonyms, and application of Boolean operators to help students understand how search strategies are systematically designed.

    \item \textbf{Criteria Chat.} The Criteria Chat assists students in defining inclusion and exclusion criteria based on their study goal or research question. This feature supports reasoning about study selection by proposing criteria and explaining how they relate to the scope and focus of the review.
\end{itemize}

Across these components, \textit{SLR Mentor} is designed to support understanding and reflection on SLR planning practices through conversational interaction. Students can engage in dialogue with the tool, building their understanding incrementally from one question to the next while practicing planning activities. Through the use of retrieval-augmented generation, responses are grounded in curated SLR guidelines and methodological documentation, ensuring that explanations remain aligned with established practices rather than relying on uncontextualized model generation. \\

\noindent \textbf{Validation.} Across the three components, participant responses suggest that \textit{SLR Mentor} is experienced primarily as a learning oriented support rather than as a tool for producing definitive planning artifacts. For the Mentor Chat (Table~\ref{tab:mentor-chat}), most participants indicate that the explanations provided are clear and generally consistent with SLR practices they already know, while perceptions of whether the answers support learning beyond providing conclusions vary across participants. For the Search String Chat (Table~\ref{tab:search-string-chat}), responses indicate that participants generally perceive the component as supporting understanding of the reasoning behind search string construction and the inclusion or exclusion of terms and operators, with stronger agreement regarding its potential usefulness for students who are learning to design search strategies. In contrast, responses for the Criteria Chat (Table~\ref{tab:criteria-chat}) show greater variability, with some participants reporting support for understanding and reflecting on inclusion and exclusion criteria, while others report lower comparability between tool-generated criteria and those defined manually.

Participants’ narratives describe \textit{SLR Mentor} as a learning-oriented support that assists with understanding the SLR process while still requiring active judgement and refinement. Several participants emphasized the clarity and adequacy of the Mentor Chat for orienting first-time reviewers, noting that responses were \textit{“clear and aligned to the response”} (P04) and that they \textit{“really help answer the question in full”} (P03). In terms of efficiency, participants reported that the tool helped structure their work when engaging with complex planning tasks. One participant described it as \textit{“generally helpful for parsing through complicated systematic literature review requirements”} (P01), while another characterized it as a \textit{“beginner-friendly scaffold”} that helped them move through the SLR workflow in a structured way (P02). The Search String Chat was described as educational rather than prescriptive, with participants highlighting that it \textit{“does a good job explaining how and why things are done”} (P01) and that it is \textit{“useful to show how one might be constructed”} (P04), while still requiring further refinement by the student. Experiences with the Criteria Chat were more mixed, but several participants highlighted its role as an initial learning aid. One participant noted that it helped illustrate \textit{“common categories used to define inclusion and exclusion criteria”} (P02), while another described it as \textit{“extremely useful”} for first time use, with the expectation that criteria would be refined iteratively (P03). Across components, participants framed the tool as supporting learning through explanation, structure, and comparison with their own work, rather than as a system that replaces methodological decision-making.

\input{tables/evaluation_mentor}

\input{tables/evaluaton_search}
\input{tables/evaluaton_criteria}

%% file: tables/evaluation_mentor.tex
\begin{table}[!ht]
\centering
\caption{Participant responses for the Mentor Chat}
\label{tab:mentor-chat}
\scriptsize
\begin{tabular}{|p{2cm}|p{3cm}|p{3cm}|p{3cm}|}
\hline
\textbf{Participant} &
The explanations provided by the Mentor Chat were clear and easy to follow. &
The information provided was consistent with SLR practices you already knew &
The answers supported learning rather than simply providing conclusions. \\
\hline
P01 & Agree & Agree & Neutral \\
\hline
P02 & Strongly agree & Strongly agree & Agree \\
\hline
P03 & Strongly agree & Agree & Strongly agree \\
\hline
P04 & Strongly agree & Strongly agree & Disagree \\
\hline
\end{tabular}
\end{table}

%% file: tables/evaluaton_search.tex
\begin{table}[!ht]
\centering
\caption{Participant responses for the Search String Chat}
\label{tab:search-string-chat}
\scriptsize
\begin{tabular}{|p{1.6cm}|p{2.4cm}|p{2.4cm}|p{2.4cm}|p{2.4cm}|}
\hline
\textbf{Participant} &
Helped me understand search string construction &
Helped me understand why terms and operators were included or excluded &
Helped me reflect on my manually constructed search string &
Could support students learning to build search strings for the first time \\
\hline
P01 & Agree & Agree & Disagree & Strongly agree \\
P02 & Agree & Agree & Strongly agree & Strongly agree \\
P03 & Strongly agree & Agree & Strongly agree & Strongly agree \\
P04 & Agree & Neutral & Agree & Strongly agree \\
\hline
\end{tabular}
\end{table}

%% file: tables/evaluaton_criteria.tex
\begin{table}[!ht]
\centering
\caption{Participant responses for the Criteria Chat}
\label{tab:criteria-chat}
\scriptsize
\begin{tabular}{|p{1.6cm}|p{2.4cm}|p{2.4cm}|p{2.4cm}|p{2.4cm}|}
\hline
\textbf{Participant} &
Helped me understand how inclusion and exclusion criteria are derived &
Criteria were comparable to my manually defined criteria &
Helped me reflect on the clarity and precision of my criteria &
Could support students defining inclusion and exclusion criteria for the first time \\
\hline
P01 & Disagree & Strongly disagree & Neutral & Neutral \\
P02 & Agree & Neutral & Agree & Agree \\
P03 & Agree & Strongly agree & Strongly agree & Strongly agree \\
P04 & Strongly agree & Agree & Agree & Strongly agree \\
\hline
\end{tabular}
\end{table}

%% file: discussion.tex
\section{Discussion} \label{sec:discussion}
Our results align with aspects of prior work on teaching systematic literature reviews in software engineering, which reports that novice researchers encounter difficulties when learning how to reason about planning decisions, which should include search strategies and inclusion and exclusion criteria~\cite{jorgensen2005teaching,castelluccia2013teaching,riaz2010experiences}. In our study, participants’ interactions with \textit{SLRMentor} indicate that the tool was mainly used to clarify concepts, revisit methodological choices, and reflect on planning decisions they had already made manually, rather than to simply follow procedural steps or adopt generated artifacts. This is compatible with earlier educational studies that emphasize the value of explanation and reflection for supporting understanding of secondary study methods~\cite{castelluccia2013teaching}. At the same time, our results extend existing work by illustrating how conversational, guideline-grounded assistance can be embedded directly into SLR planning activities, providing learners with on-demand explanations and opportunities for iterative sense-making during practice, rather than relying solely on instructor-led feedback or static training materials. Answering our research question (\textbf{RQ.} \textit{How can a conversational assistant support learning SLR processes for software engineering researchers?}), our results indicate that a conversational assistant can support learning by making methodological reasoning more explicit during planning activities and by enabling comparison between tool-supported guidance and researchers’ own decisions. Participants’ accounts also point to clear limits of this support, as meaningful use of the tool required critical judgment and domain knowledge, suggesting that conversational assistants may complement, but do not replace, established approaches to teaching systematic literature reviews in software engineering.

\subsection{Implications for Research and Education}
The results suggest implications for both research and education in software engineering. From a research perspective, the findings indicate that conversational assistants grounded in established guidelines can support learning by making methodological reasoning explicit during systematic literature review planning, rather than by producing finalized artifacts. This points to the need for further investigation of such tools as supports for reflection, comparison, and sense making, including how they influence the development of independent methodological judgment over time and across levels of research experience. From an educational perspective, the results indicate that \textit{SLRMentor} can function as a complementary scaffold that supports understanding and orientation during SLR activities while still requiring active student judgment and refinement. This suggests that conversational assistants may be integrated into teaching as on-demand supports that reinforce methodological concepts and structure practice, provided they are positioned as aids for learning and reflection rather than as replacements for instruction, supervision, or critical decision making.

\subsection{Threats to Validity}

Following established empirical standards in software engineering research~\cite{ralph2020empiricalstandards}, threats to validity should be acknowledged. The pilot validation of \textit{SLRMentor} involved only four graduate students, which limits the strength of any generalizable claims. However, this sample represents half of the students enrolled in the target graduate course, and the validation should therefore be interpreted in light of the characteristics and scale of the educational setting. The study does not aim to provide statistical validation, but rather an initial assessment of how the tool is experienced as a learning support. As such, the results should be understood as indicative and exploratory, intended to inform the design of a larger-scale validation and subsequent tool refinement. Further evaluations with broader and more diverse student populations are required to assess transferability and to strengthen the empirical basis of the findings.

%% file: conclusions.tex
\section{Conclusions and Future Work} \label{sec:conclusions}
This paper presented \textit{SLR Mentor}, a conversational assistant designed to support the teaching and learning of systematic literature review planning in software engineering. The tool addresses key planning activities, including search string construction, reasoning about inclusion and exclusion criteria, and general methodological guidance, with responses grounded in established SLR guidelines through retrieval augmented generation. The results indicate that \textit{SLR Mentor} is experienced primarily as a learning-oriented support that assists novice researchers in understanding and reflecting on planning decisions, rather than as a system for producing finalized review artifacts. In this way, the tool lowers initial barriers to engagement with SLR methods while maintaining the need for active judgment and methodological responsibility. Future work will focus on examining learning outcomes through larger-scale empirical studies, expanding the curated retrieval corpus, and extending support to later stages of the SLR process, including study selection, data extraction, quality assessment, and synthesis. Extending the assistant to these stages would enable learners to engage with methodological reasoning beyond planning, supporting consistent decision-making across the full review lifecycle.

%% file: bib.bib
@inproceedings{kitchenham2004ebse,
  author    = {Kitchenham, Barbara A. and Dyb{\aa}, Tore and J{\o}rgensen, Magne},
  title     = {Evidence-Based Software Engineering},
  booktitle = {Proceedings of the 26th International Conference on Software Engineering (ICSE 2004)},
  publisher = {IEEE},
  year      = {2004},
  pages     = {273--281}
}

@article{kitchenham2009slr,
  author  = {Kitchenham, Barbara and Brereton, O. Pearl and Budgen, David and Turner, Mark and Bailey, John and Linkman, Stephen},
  title   = {Systematic Literature Reviews in Software Engineering -- A Systematic Literature Review},
  journal = {Information and Software Technology},
  volume  = {51},
  number  = {1},
  pages   = {7--15},
  year    = {2009},
  url     = {https://www.sciencedirect.com/science/article/pii/S0950584908001390}
}

@inproceedings{petersen2008sms,
  author    = {Petersen, Kai and Feldt, Robert and Mujtaba, Shahid and Mattsson, Michael},
  title     = {Systematic Mapping Studies in Software Engineering},
  booktitle = {Proceedings of the 12th International Conference on Evaluation and Assessment in Software Engineering (EASE)},
  publisher = {BCS Learning \& Development},
  year      = {2008}
}

@inproceedings{santos2014smsreviews,
  author    = {Santos, Ronnie E. and de Magalh{\~a}es, Cleyton V. and da Silva, Fabio Q. B.},
  title     = {The Use of Systematic Reviews in Evidence Based Software Engineering: A Systematic Mapping Study},
  booktitle = {Proceedings of the 8th ACM/IEEE International Symposium on Empirical Software Engineering and Measurement},
  year      = {2014},
  pages     = {1--4}
}

@article{wohlin2020updatesearch,
  author  = {Wohlin, Claes and Mendes, Emilia and Felizardo, Katia R. and Kalinowski, Marcos},
  title   = {Guidelines for the Search Strategy to Update Systematic Literature Reviews in Software Engineering},
  journal = {Information and Software Technology},
  volume  = {127},
  pages   = {106366},
  year    = {2020}
}

@inproceedings{santos2013motivation,
  author    = {Santos, Ronnie E. and da Silva, Fabio Q. B.},
  title     = {Motivation to Perform Systematic Reviews and Their Impact on Software Engineering Practice},
  booktitle = {Proceedings of the ACM/IEEE International Symposium on Empirical Software Engineering and Measurement},
  publisher = {IEEE},
  year      = {2013},
  pages     = {292--295}
}

@inproceedings{zhang2010searching,
  author    = {Zhang, He and Babar, Muhammad Ali},
  title     = {On Searching Relevant Studies in Software Engineering},
  booktitle = {Proceedings of the 14th International Conference on Evaluation and Assessment in Software Engineering (EASE)},
  year      = {2010},
  pages     = {111--120}
}

@inproceedings{felizardo2024chatgptslr,
  author    = {Felizardo, K{\'a}tia Romero and Lima, M{\'a}rcia Sampaio and Deizepe, Anderson and Conte, Tayana Uch{\^o}a and Steinmacher, Igor},
  title     = {ChatGPT Application in Systematic Literature Reviews in Software Engineering: An Evaluation of Its Accuracy to Support the Selection Activity},
  booktitle = {Proceedings of the 18th ACM/IEEE International Symposium on Empirical Software Engineering and Measurement (ESEM '24)},
  year      = {2024},
  pages     = {25--36},
  doi       = {10.1145/3674805.3686666}
}

@article{delatorre2023ai,
  author  = {de la Torre-L{\'o}pez, Jos{\'e} and Ram{\'i}rez, Antonio and Romero, Jos{\'e} R.},
  title   = {Artificial Intelligence to Automate the Systematic Review of Scientific Literature},
  journal = {Computing},
  volume  = {105},
  number  = {10},
  pages   = {2171--2194},
  year    = {2023}
}

@article{bolanos2024ailitreviews,
  author  = {Bola{\~n}os, Francisco and Salatino, Angelo and Osborne, Francesco and Motta, Enrico},
  title   = {Artificial Intelligence for Literature Reviews: Opportunities and Challenges},
  journal = {Artificial Intelligence Review},
  volume  = {57},
  number  = {10},
  pages   = {259},
  year    = {2024}
}

@article{ralph2020empiricalstandards,
  author  = {Ralph, Paul and Ali, Nauman bin and Baltes, Sebastian and Bianculli, Domenico and Diaz, Jessica and Dittrich, Yvonne and Ernst, Neil and Felderer, Michael and Feldt, Robert and Filieri, Antonio and others},
  title   = {Empirical Standards for Software Engineering Research},
  journal = {arXiv preprint arXiv:2010.03525},
  year    = {2020},
  url     = {https://arxiv.org/abs/2010.03525}
}

@inproceedings{wessel2022bots,
  author    = {Wessel, Mairieli and Gerosa, Marco A. and Shihab, Emad},
  title     = {Software Bots in Software Engineering: Benefits and Challenges},
  booktitle = {Proceedings of the 19th International Conference on Mining Software Repositories (MSR '22)},
  year      = {2022},
  pages     = {724--725}
}

@inproceedings{jorgensen2005teaching,
  title={Teaching evidence-based software engineering to university students},
  author={Jorgensen, Magne and Dyba, Tore and Kitchenham, Barbara},
  booktitle={11th IEEE International Software Metrics Symposium (METRICS'05)},
  pages={8--pp},
  year={2005},
  organization={IEEE}
}

@article{castelluccia2013teaching,
  title={Teaching evidence-based software engineering: learning by a collaborative mapping study of open source software},
  author={Castelluccia, Daniela and Visaggio, Giuseppe},
  journal={ACM SIGSOFT Software Engineering Notes},
  volume={38},
  number={6},
  pages={1--4},
  year={2013},
  publisher={ACM New York, NY, USA}
}

@article{pizard2021training,
  title={Training students in evidence-based software engineering and systematic reviews: a systematic review and empirical study},
  author={Pizard, Sebasti{\'a}n and Acerenza, Fernando and Otegui, Ximena and Moreno, Silvana and Vallespir, Diego and Kitchenham, Barbara},
  journal={Empirical Software Engineering},
  volume={26},
  number={3},
  pages={50},
  year={2021},
  publisher={Springer}
}

@inproceedings{riaz2010experiences,
  title={Experiences conducting systematic reviews from novices’ perspective},
  author={Riaz, Mehwish and Sulayman, Muhammad and Salleh, Norsaremah and Mendes, Emilia},
  booktitle={14th International Conference on Evaluation and Assessment in Software Engineering (EASE)},
  year={2010},
  organization={BCS Learning \& Development}
}

@article{fernandez2024exploring,
  title={Exploring the Frontier of Software Engineering Education with Chatbots},
  author={Fernandez-y-Fernandez, CA and S{\'a}nchez-Soto, E and Cisnero, JR Aguilar and Ju{\'a}rez-Ram{\'\i}rez, R},
  journal={Programming and Computer Software},
  volume={50},
  number={8},
  pages={796--815},
  year={2024},
  publisher={Springer}
}

@article{sengul2024software,
  title={Software engineering education in the era of conversational AI: current trends and future directions},
  author={Sengul, Cigdem and Neykova, Rumyana and Destefanis, Giuseppe},
  journal={Frontiers in Artificial Intelligence},
  volume={7},
  pages={1436350},
  year={2024},
  publisher={Frontiers Media SA}
}

@inproceedings{chan2023potential,
  title={The potential role of AI-based Chatbots in Engineering Education. Experiences from a teaching perspective},
  author={Chan, Miguel Morales and Amado-Salvatierra, Hector R and Hernandez-Rizzardini, Rocael and De La Roca, M{\'o}nica},
  booktitle={2023 IEEE frontiers in education conference (FIE)},
  pages={1--5},
  year={2023},
  organization={IEEE}
}

@inproceedings{cunningham2019review,
  title={A review of chatbots in education: practical steps forward},
  author={Cunningham-Nelson, Sam and Boles, Wageeh and Trouton, Luke and Margerison, Emily},
  booktitle={30th annual conference for the australasian association for engineering education (AAEE 2019): educators becoming agents of change: innovate, integrate, motivate},
  pages={299--306},
  year={2019},
  organization={Engineers Australia}
}

@inproceedings{khan2025integrating,
  title={Integrating llms in software engineering education: Motivators, demotivators, and a roadmap towards a framework for finnish higher education institutes},
  author={Khan, Maryam and Akbar, Muhammad Azeem and Kasurinen, Jussi},
  booktitle={Proceedings of the 2025 29th International Conference on Evaluation and Assessment in Software Engineering Companion},
  pages={182--191},
  year={2025}
}

@article{swacha2025retrieval,
  title={Retrieval-Augmented Generation (RAG) Chatbots for Education: A Survey of Applications},
  author={Swacha, Jakub and Gracel, Micha{\l}},
  journal={Applied Sciences},
  volume={15},
  number={8},
  pages={4234},
  year={2025},
  publisher={MDPI}
}

@inproceedings{fan2024survey,
  title={A survey on rag meeting llms: Towards retrieval-augmented large language models},
  author={Fan, Wenqi and Ding, Yujuan and Ning, Liangbo and Wang, Shijie and Li, Hengyun and Yin, Dawei and Chua, Tat-Seng and Li, Qing},
  booktitle={Proceedings of the 30th ACM SIGKDD conference on knowledge discovery and data mining},
  pages={6491--6501},
  year={2024}
}
